\documentstyle[amsmath,amsfonts,epsfig,12pt]{article}

\newcommand {\slsh} [1] {\not{\hbox{\kern-2pt${#1}$}}}

\def\drawbox#1#2{\hrule height#2pt
         \hbox{\vrule width#2pt height#1pt \kern#1pt
               \vrule width#2pt}
               \hrule height#2pt}

\def\Asym#1#2{\vcenter{\vbox{\drawbox{#1}{#2}
               \kern-#2pt       
               \drawbox{#1}{#2}}}}

\newcommand {\beq} {\begin{equation}}
\newcommand {\eeq} {\end{equation}}
  \newcommand {\ber}{\begin{eqnarray*}}
  \newcommand {\eer} {\end{eqnarray*}}
\newcommand {\bea}{\begin{eqnarray}}
  \newcommand {\eea} {\end{eqnarray}}

\newcommand{\Dslash}{\,{\raise.15ex\hbox{/}\mkern-12mu D}}

\begin{document}


\begin{titlepage}
\rightline{IFUP-TH/2011-17}
\rightline{CERN-PH-TH/2011-217}
\begin{center}
\vspace{0.1in}
\large{\bf $k$-String Tension from Eguchi-Kawai Reduction}\\
\vspace{0.4in}
\large{Adi Armoni $^{a}$, Daniele Dorigoni $^{b,c}$, Gabriele Veneziano $^{d,e}$}\\
\vspace{0.1in}
\small{\texttt{a.armoni@swan.ac.uk, d.dorigoni@sns.it, gabriele.veneziano@cern.ch}}\\
\vspace{0.2in}

\large{\emph {$^{a}$Department of Physics, Swansea University}}\\ 
\large{\emph {Singleton Park, Swansea, SA2 8PP, UK}}\\
\vspace{0.3in}

\large{\emph {$^{b}$Scuola Normale Superiore,}}\\
\large{\emph {Piazza dei Cavalieri 7, 56126 Pisa, Italy }}\\
\vspace{0.1in}
\large{\emph {$^{c}$Istituto Nazionale di Fisica Nucleare, Sezione di Pisa}} \\
\large{\emph {Via Buonarroti, 2, Ed. C, 56127 Pisa,  Italy}}\\
\vspace{0.3in}

\large{\emph {$^{d}$Coll\`ege de France,}}\\
\large{\emph {11 Place M. Berthelot, 75005 Paris, France}}\\ 
\vspace{0.1in}
\large{\emph {$^{e}$CERN, Theory Unit, Physics Department,}} \\
\large{\emph {CH-1211 Geneva 23, Switzerland}}\\
\vspace{0.3in}

\end{center}

\abstract{We consider three-dimensional $SU(N)$ gauge theories with massless Dirac or Majorana fermions in the adjoint representation. We use the Eguchi-Kawai volume reduction to two-dimensions to calculate the tension $\sigma _ k$ of the $k$-string in such theories. Under some assumptions whose validity we discuss, we derive the previously conjectured sine formula, $\sigma _ k \sim N \sin ( \pi {k\over N})$.}

\end{titlepage}

\section{Introduction}

 The dynamics of QCD in the non-perturbative regime is extremely complex and difficult. In order to simplify the problem 't Hooft  suggested a long time ago to replace the gauge group $SU(3)$ of QCD by an $SU(N)$ group and to take the 
large-$N$ limit, while keeping $g^2 N$ fixed \cite{'tHooft:1973jz}. In this limit the theory is expected to be controlled by a classical master field. Yet, despite the expectation of a huge simplification, very few analytical results were achieved by using the large-$N$ approximation.

 An important example of these is the Eguchi-Kawai volume reduction \cite{Eguchi:1982nm}. It was argued that when a planar theory is compactified on a circle, quantities that are neutral under the center of the gauge group should not depend on the radius of the circle and hence calculations of neutral quantities by a reduction to an arbitrarily small volume should be possible. Shortly after  the seminal paper by Eguchi and Kawai, it was shown that a necessary condition for the reduction is that the $Z_N$ center of the gauge group is not spontaneously broken, namely that the theory is confining at arbitrarily small volumes \cite{Bhanot:1982sh}. Unfortunately the pure $SU(N)$ Yang-Mills theory undergoes a confinement/deconfinement phase transition at a certain critical radius and hence a reduction to a radius below the critical radius is not allowed. 

The idea of volume reduction was resurrected recently by Kovtun, Unsal and Yaffe \cite{Kovtun:2007py}. They argued that the Eguchi-Kawai reduction can be used for QCD-like theories with fermions that transform in the {\it adjoint} representation of the gauge group. They showed that if the adjoint fermions obey periodic boundary conditions (hence opposite boundary conditions w.r.t. the finite temperature case), the theory is confining at an arbitrarily small radius and the condition for volume reduction is satisfied. Following their work several papers \cite{adjoint-EK} supporting the validity of volume reduction for theories with adjoint fermions were published.   This observation becomes particularly interesting when combined with another  intriguing outcome of the large-$N$ limit:  a set of equivalences amongst
seemingly distinct gauge theories \cite{Armoni:2004uu,Kovtun:2004bz}. In particular, a sector of QCD observables has a counterpart in the theory with adjoint fermions. In principle this could open the way to computing physical QCD quantities in the large-$N$ limit through small volume calculations.

In this paper we consider three-dimensional gauge theories with adjoint fermions and use volume-reduction to two-dimensions in order to calculate the $k$-string tension. The $k$-string can be thought of as a bound state of $k$ fundamental QCD-strings. Its tension has been vastly discussed in the literature \cite{kstring}. In particular, most analytic approaches in the case of theories with adjoint fermions suggest that the $k$-string tension is proportional to $\sin (\pi {k\over N})$ \cite{sin}.

Our derivation of the string tension is similar to the corresponding one in the 2d Abelian (massive Schwinger model) \cite{Coleman:1975pw} and the non-Abelian \cite{Gross:1995bp,Armoni:1997ki} cases. We rely on the fact that in two dimensions an external charge at spatial infinity is equivalent to a $\theta$-term \cite{Witten:1978ka}. The main result of this paper is 
\beq
\sigma_k  \sim N \sin \left ( \pi {k\over N} \right ) .
\eeq

The paper is organized as follows: in section 2 we review the derivation of the string tension in the Abelian case. In section 3 we derive the $k$-string tension for the dimensionally reduced theory. In section 4 we discuss the validity of our derivation. Section 5 is devoted to a summary. In the appendix we discuss how the derivation of section 3 can be repeated in the case of a 3d theory with a single Majorana adjoint fermion.

\section{A reminder of the string tension in the massive Schwinger model}

Consider two-dimensional QED with one massive Dirac fermion whose electric charge (in units of the elementary charge)
is $Q$. The action is
\beq
S=\int d^2 x \, \left ( -{1\over 4e^2} F_{\mu \nu} ^2 + \bar \Psi i \slsh \partial \Psi - m\bar \Psi \Psi +  Q  A_{\mu} \bar \Psi \gamma ^{\mu} \Psi  
\right ) \, ,
\eeq
where $e$ is the gauge coupling. It is convenient to bosonize the fermion and to use the gauge $A_1 =0$. Denoting the boson by $\psi$ the resulting action takes the form
\beq
S=\int d^2 x \, \left ( {1\over 2e^2} (\partial _1 A_0)^2 + {1\over 2} (\partial _\mu \psi)^2 + m \mu \cos (2\sqrt \pi \psi) +  {Q\over \sqrt \pi} A_0 \partial _1 \psi  \right  ) \, , 
\eeq
where $\mu=e {\exp \gamma \over 2 \pi ^{3/2}}$. 
In the massless limit diagonalization of the quadratic terms reveals the existence of a free massive boson with
$m^2 = \frac{e^2 Q^2}{\pi}$. When $m \ll e$, namely when the mass term is considered as a small perturbation, the vacuum configuration corresponds to $\psi =0$ for $m > 0$ and to $2\sqrt \pi \psi = \pi$ for $m <0$.
The vacuum energy is negative and given  by $- |m| \mu$.

Let us add a heavy electron-positron pair of charge $\pm Q'$ at $x=\pm L$ having in mind to take the limit  $L \rightarrow \infty$ at the end. This means adding the term
\beq
S' = \int d^2 x \,  Q' A_0 \left (\delta (x-L) - \delta (x+L) \right ) 
\eeq
to the previous action. We can integrate out the non-dynamical field $A_0$,
\beq
\partial _1 A_0 = e^2 \left ( {Q \over \sqrt \pi} \psi + Q'( \theta (x-L) - \theta (x+L) ) \right )\; ,
\eeq
to arrive at the following action:
\beq
S=\int d^2 x \,  \left ( {1\over 2} (\partial _\mu \psi)^2  +  m \mu \cos (2\sqrt \pi \psi)  - \frac{e^2}{2} ({Q\over \sqrt \pi}\psi + Q' \Theta)^2 \right ) \, ,
\eeq
where $\Theta \equiv \theta (x-L) - \theta (x+L)$.

The corresponding Hamiltonian of the theory is given by
\beq
H=\int dx \, \left ( {1\over 2} \Pi_{\psi} ^2 + {1\over 2} (\partial _1 \psi)^2
 -  m \mu \cos (2\sqrt \pi \psi) +  \frac{e^2}{2} ({Q\over \sqrt \pi}\psi + Q' \Theta)^2 \right )  \, . \label{Abel-hamilton}
\eeq
Let us restrict ourselves to the limit $0 < m\mu /e ^2 \ll 1$. In this limit  the solution of the equation of motion for $\psi$ is 
\beq
 \psi = - \sqrt \pi {Q' \Theta \over Q} + {\cal O}(\frac{m \mu}{ e^2})   \,. \label{Abel-sol}
\eeq
The vacuum energy is given by substituting \eqref{Abel-sol} in \eqref{Abel-hamilton}:
\beq
E = H-H_0 =  m\mu  \left(1-\cos (2\pi {Q' \over Q}) \right )2L \, ,
\eeq
where $H_0$ is the Hamiltonian of the theory without the external charge. After dealing similarly with the case of a small negative mass we find that  the string tension is 
\beq
 \sigma = |m| \mu \left (1-\cos (2\pi {Q' \over Q}) \right ) \,. \label{Astring}
\eeq
The  tension \eqref{Astring} vanishes if $Q'$ is a multiple of $Q$ and, interestingly, also when the quark mass is zero (corresponding to total charge screening).

\section{The string tension in three-dimensional adjoint QCD}

Consider three-dimensional adjoint QCD with one Dirac fermion, based on a gauge group $U(N)$:
\begin{equation}
 {\mathcal{L}} = \mbox{Tr} \left( -\frac{1}{2 g_3^2}  F^{mn} F_{m n} + i\bar{\psi} \Dslash \psi \right) ~~;~~ m,n=0,1,2 \,,
\end{equation}
and compactified on ${\mathbb{R}}^2 \times S^1$.
 By imposing periodic boundary conditions on both the gauge and fermionic fields,
the full action, including the Kaluza-Klein modes, is given by:
\bea
& & \sum_{q\in {\mathbb{Z}}} \mbox{Tr} \left( 
-\frac{1}{2g^2} F^{\mu\nu}_{-q} F_{q\,\mu\nu}+(D^\mu \phi)_{-q}(D_\mu \phi)_{q}-\frac{2}{gR} A^\mu_{-q} (D_\mu \phi)_{q}
+\frac{q^2}{g^2 R^2} A^\mu_{-q} A_{q\,\mu}+ \nonumber
\right.\\
& & \left. +i{\bar{\psi}}_{-q} \gamma^\mu (D_\mu \psi)_{q}-\frac{q}{R} {\bar{\psi}}_{-q} \gamma^2 \psi_q
-g\sum_{q_1} {\bar{\psi}}_{-q}\gamma^2 [\phi_{q_1}.\psi_{q-q1}] \right) \, ,
\eea
where $g = g_3/\sqrt{R}$ is the 2-dimensional gauge coupling, the ``covariant'' derivative on a generic field $\varphi$ is given by $(D^\mu \varphi)_{q}=\partial^\mu \varphi_q+i \sum_{q_1} [A_{q_1}^\mu,\varphi_{q-q_1}]$ with $\mu,\nu=0,1$, 
and we introduced the tower of scalars $A_{2}/g= \sum_{q\in{\mathbb{Z}}} \phi_q\, e^{i q y/R}$.

Usual Kaluza-Klein reduction gives a tower of states whose only massless fields are those related to the ``zero"   modes with $q=0$.  All the others acquire a mass of order $q/R$, where $R$ is the radius of $S^1$. This is no longer  true  when the scalars of the theory acquire v.e.v.s, since these modify the mass spectra through Yukawa and gauge interactions and, as a result,  the different mass levels get shifted and split.

V.e.v.s for the scalars arise quite naturally. Let us denote the $q=0$ component of the gauge field along the circle by $\phi$ and study its vev.
 The vacuum configuration, 
obtained by a minimization of the effective action for the Polyakov loop ${\rm Tr}\, \exp (i \int _0 ^{2\pi} \phi)$, yields 
\cite{Kovtun:2004bz}:
\beq 
\langle \phi \rangle _{nm}= v_n \delta _{nm} \equiv {(n - (N+1)/2) \over N} \delta _{nm} \label{vevs}
\eeq
and hence the center-symmetry is not  spontaneously broken. 

Let us focus for the moment on the zero modes only, namely let us carry out a dimensional reduction down to two dimensions. 
As just mentioned,  actually this is not necessarily justified. 
The correct procedure is to to keep the lowest mass modes, which are not necessarily the $q=0$ modes: 
the lightest modes may include KK momentum or winding. We will discuss this important issue in section 4.

The reduced action is
\beq 
S=\int d^2 x \, {\rm Tr} \left ( -{1\over 2g^2} F_{\mu \nu}^2 + (D_\mu \phi)^2 + \bar \Psi i \slsh D \Psi + ig [\phi, \bar \Psi] \gamma ^3 \Psi \right ) \, ,
\eeq
where the matrix $\gamma_3=\mbox{diag} (1,-1)$ is the chiral matrix in $2d$ which is equal to $-i\gamma^2$ in $3d$ and the Yukawa coupling  is $g\sim {g_{3}\over \sqrt{R}}$. 
Let us choose again the gauge $A_1=0$. In the presence of the v.e.v. \eqref{vevs} the gauge symmetry is broken to $U(1)^N$, with the massless gauge fields given by 
\beq
(A_0)_{nm} \equiv A_n \delta _{nm} \, .
\eeq

Let us focus on the gauge and Yukawa interactions, since these terms in the action will determine the 2d string tension. In the presence of the v.e.v. the terms
\bea
 {\rm Tr}\,  i [\phi, \bar \Psi]\gamma ^3 \Psi~~;~~
 {\rm Tr} \,  [A_0, \bar \Psi] \gamma ^0 \Psi
\eea
become, respectively: 
\bea
  i(v_n - v_m) \bar \Psi _{nm} \gamma ^3 \Psi _{mn} ~~ ; ~~
 (A_n - A_m) \bar \Psi _{nm} \gamma ^0 \Psi _{mn} \, .
\eea
The action for the $U(1)^N$ gauge theory, excluding the kinetic terms for the scalar and  fermi fields --which do not play any role in the calculation of the string tension-- is:
\bea
& & S=\int d^2 x \, \left \{ \sum _n {1\over 2g^2} (\partial _1 A_n)^2 \right . \nonumber \\
& &
\left . + m \sum _{mn}i(v_n - v_m) \bar \Psi _{nm} \gamma ^3 \Psi _{mn}+  \sum_{mn}  (A_n - A_m) \bar \Psi _{nm} \gamma ^0 \Psi _{mn} \right \} \,.
\eea
As the derivation of the string tension involves terms which are not invariant under chiral rotation, we should also incorporate the chiral anomaly in the action. In analogy with the well-known case of QCD \cite{QCDLeff}, the effective action can be written in the form:
\bea
& & S_{eff}=\int d^2 x \, \left \{ \sum _n {1\over 2g^2} F_n^2
 + m \sum _{mn}i(v_n - v_m) \bar \Psi _{nm} \gamma ^3 \Psi _{mn} +
 \right . \\
& &
\left .  \sum_{mn}  (A_n - A_m) \bar \Psi _{nm} \gamma ^0 \Psi _{mn}+
 \sum_n  F_n \left[\frac{-i}{8\pi}\sum_{m} \ln \left(
\frac{\Psi^{\ast}_{L\,nm}\Psi_{R\,mn}\,\,\Psi^{\ast}_{R\, mn}\Psi_{L\,nm}}{\Psi^{\ast}_{L\, mn}\Psi_{R\,nm}\,\,
 \Psi^{\ast}_{R\, nm}\Psi_{L\,mn}} \right) \right]
 \right \} \nonumber  \,,
\eea
where $F_n = \partial _1 A_n $. 

One can easily check that, under a chiral transformation in the $a^{th}$ direction of the Cartan subalgebra,
\bea
  & & \delta \Psi_{L\, ij} = i\alpha\delta_{ia}\Psi_{L\, aj} -i\alpha \Psi_{L\, ia}\delta_{aj} + O(\alpha^2)\\
  & &\delta \Psi^\ast_{L\, ij} = i\alpha\delta_{ia}\Psi^\ast_{L\, aj} -i\alpha \Psi^\ast_{L\, ia}\delta_{aj} + O(\alpha^2)\\
 & &\delta \Psi_{R\,ij}=\delta \Psi_{R\,ij}^\ast=0 \, ,
\eea
the effective action transforms as it should:
 \begin{equation*}
 \delta_a {\mathcal{L}}_{eff} = \frac{1}{2\pi}(\alpha N F^a -\alpha \sum_{i=1}^N F^i)\, .
\end{equation*}
Clearly the overall $U(1)$ is not anomalous since  $\sum_{a=1}^N \delta_a {\mathcal{L}}=0$. 
Consequently, we have selected indeed an $SU(N)$, rather than a $U(N)$, anomaly.

Let us add now a source of the form 
\bea
& & S' = \int d^2 x \, \left( {k\over 2}  A^a - {k\over 2} A^b
\right)\left (\delta (x-L) - \delta (x+L) \right ) = \\
& & = -\int d^2 x \, \left( {k\over 2} F_a - {k\over 2}F_b \right) \Theta \, ,
\eea
that corresponds to $k$ units of fundamental charge placed at the end of the interval and points along the $SU(N)$ Cartan subalgebra, namely
 $\vec{k}=(0,...,0,k,0,...,0,-k,0,...)$.
 
In the presence of the external charge the relevant part of the effective action is
\bea
& & S_{eff}=\int d^2 x \, \left \{ \sum _n {1\over 2g^2} F_n^2
 + m \sum _{mn}i(v_n - v_m) \left(
\Psi^\ast_{L\,nm}\Psi_{R\, mn}-\Psi^\ast_{R\,nm}\Psi_{L\, mn}
\right)  + \right . \nonumber \\
& &
\left .
 \sum_n  F_n \left[\frac{-i}{8\pi}\sum_{m} \ln \left(
\frac{\Psi^{\ast}_{L\,nm}\Psi_{R\,mn}\,\,\Psi^{\ast}_{R\, mn}\Psi_{L\,nm}}{\Psi^{\ast}_{L\, mn}\Psi_{R\,nm}\,\,
 \Psi^{\ast}_{R\, nm}\Psi_{L\,mn}} \right) -{k\over 2} \delta ^a_n + {k \over 2} \delta ^b _n \right]
 \right \}  \,.
\eea
In order to write the effective action in terms of bosonic fields let us introduce
\bea
 \Psi^{\ast}_{L\, nm} \Psi_{R\, mn} = \mu \exp (i \psi _{mn})~~ ; ~~
 \Psi^{\ast}_{R\, mn} \Psi_{L\,nm}= \mu \exp (-i \psi _{nm}) \,.
\eea
The bosonized action takes the form
\bea
& & S_{eff}=\int d^2 x \, \left \{ \sum _n {1\over 2g^2} F_n^2
 - 2 m\mu \sum _{mn}(v_n - v_m)  \sin \psi _{mn}  + \right . \nonumber  \\
& &\label{Seff}
\left .
 \sum_n  F_n \left[\frac{-i}{8\pi}\sum_{m} \ln 
\exp \left(2i (\psi _{mn} - \psi _{nm}) \right)  -{k\over 2} \delta _n ^a + {k \over 2} \delta _n ^b \right]
 \right \}  \,.
\eea
Integrating out $F_n$ the Hamiltonian density becomes:
\bea
 H  &=& 2 m\mu \sum _{mn}(v_n - v_m)  \sin \psi _{mn} \nonumber \\
 &+&
g^2 \sum _n  \left[\frac{1}{4\pi}\sum_{m}  (\psi _{mn} - \psi _{nm}) + {2 \pi q_n \over 8\pi}  -{k\over 2} \delta ^a _n + {k \over 2} \delta ^b _n \right]^2 \, , \label{hamilton}
\eea
where the multivaluedness of the logarithm is encoded into $q_n \in \mathbb{Z}$.

As for the Schwinger model we should now solve the equations of motion and evaluate $H$ on-shell. The problem at hand is greatly simplified in the limit $m/gN \ll 1$. The full set of equations of motion, which also includes
the equation of motion for the scalar,
\beq
 (D_\mu D^\mu \phi) _n = 2 m \sum_{l\neq n} \sin \psi_{nl} \, , \label{scalar}
\eeq
 is solved by:
\bea
 & &  \psi_{am}= \pi + {\pi k \over N}\,,\,\, M_{ma}<0\,, \label{SolFirst} \\
 & &  \psi_{am} = {\pi k \over N} \,,\,\,M_{ma}>0\,, \\
 & &  \psi_{bm}= \pi - {\pi k \over N}\,,\,\, M_{mb}>0\,, \\
 & &  \psi_{bm} = - {\pi k \over N} \,,\,\,M_{mb}<0\,, \\
 & &  \psi_{mn}=0\,,\,\,m,n\neq a,b\,, \label{SolLast}\\
 & &  \psi_{mn} = - \psi _{nm} \,,
\eea 
where $M_{mn}$ is the mass coefficient multiplying $\sin \psi_{nm}$ (we will discuss in the next section the importance of KK modes
and how they modify this mass matrix), while the scalar fluctuations are set to zero: $\phi=0$.

The integers $q_i$ are chosen in such a way that the quadratic term, which is the dominant contribution to the energy, vanishes on-shell and
the only non-zero piece comes from the sum of the mass terms. 

Note that an exact solution for the equation of motion for the scalar \eqref{scalar} requires $b-a = N/2$.
Substituting the solution in the Hamiltonian we find the energy density 
\beq
\langle H \rangle  = 2 m\mu \sum _{mn} (v_n - v_m) \sin \psi _{mn}= {8 m\mu\over N} \sum _{n=0}^N |n - {N \over 2}| \sin \left ( \pi {k\over N}
 \right ) \, . \label{sigmak1}
\eeq
Hence
\beq
\sigma_k \sim  m\mu N \sin \left ( \pi {k\over N}
 \right ) \, . \label{sigmak2}
\eeq

Had we started directly with a two-dimensional theory, $m$ would have been an arbitrary mass parameter, thus depending on its sign 
we should have found either the solution just proposed (when $m>0$) or another solution shifted by $\pi$ when $m<0$, precisely as it happens
in the Abelian case.
Independently on the sign of $m$ the string tension will always be $\sigma_k \sim  \vert m\vert \mu N \sin \left ( \pi {k\over N}\right) > 0$.

\section{Comments on the validity of the derivation}

The main result of our paper is the string tension \eqref{sigmak2}. In this section we discuss the assumptions that were made in our derivation and the validity of the result.

Similarly to the Abelian case (the massive Schwinger model that we reviewed in section 2), one needs to assume that the mass term is smaller than the gauge-interaction term. Since the mass term is ${\cal O}(N^2)$, while the gauge-interaction term is ${\cal O}(N^3)$, the condition is
\beq
 m \mu N^2 \ll g^2 N^3 \,. \label{condition1}
\eeq
By using $m \sim {1\over R}$, $\mu \sim g$ and the relation between the 3d gauge coupling and the 2d gauge coupling $g^2 R = g_3 ^2$, we can rewrite the above 
condition as
\beq
\lambda_3 RN \gg 1 \, ,
\eeq
where $\lambda _3$ is the 3d 't Hooft coupling. We now encounter \cite{Unsal:2010qh} the following difficulty: the masses of the KK modes and the lightest W-bosons are both 

\beq
M_{KK}\sim M_W \sim {1 \over RN} \, ,
\eeq
hence the condition that the mass term is a small perturbation implies that the W-bosons mass (in units of the 3d 't Hooft coupling) goes to zero, namely
\beq
{M_W \over \lambda_3} \sim {1\over \lambda_3 RN} \ll 1 \,. \label{condition2}
\eeq
 If the W-bosons become massless, our assumption that the dynamics is controlled by the Cartan sub-algebra degrees of freedom may be invalid. Note that \eqref{condition2} does not invalidate our derivation of the string tension within the 2d framework: within 2d we can always assume a small mass term. The problem is that the Eguchi-Kawai procedure requires a full non-Abelian dynamics and that an ``Abelianization'' of the problem may not be trusted. We will return to this issue in the summary section. 

Another important issue is the inclusion of KK modes. The mass of a generic mode that couples to the fermion bi-linear is
\beq
M^l _{mn} = {1\over R}(l + w_m - w_n + {m-n \over N}) \label{KKmodes}
\eeq
where $l$ is the KK momentum and $w$'s are integers that correspond to winding (the Polyakov loop is defined only mod $N$).
It means that when $(m-n) \sim {\cal O}(N)$ there is no separation between the lowest KK modes and higher modes. 
For this reason we should, in principle, consider all the modes and dimensional reduction is invalid. In particular the lowest $N$ fermionic masses are in the range $M^l_{mn}R N \in (-N/2,...,N/2)$ and they include modes with $w=0,\pm 1$. These are the $\psi _{ij}$ modes that we used in our derivation, {\it not} the zero modes. The rest of the modes were set to zero.

Returning to the issue of the W's, we can require that their mass $M_W\sim {1\over RN}$ be fixed in terms of $\lambda_3$ which means $1/R = r \lambda_3 N $ with $r$ a dimensionless number.
The expansion parameter for our previous solution $\epsilon = {m \mu\over g^2 N} $ can be rewritten using all the above formulas and it is precisely
$\epsilon=r$. The proposed solution admits a series expansion in power of $\epsilon$ and at first order we obtained eq.(\ref{sigmak2}) which
can be rewritten in the more suggestive form:
\begin{equation}
 \sigma_k \sim \lambda_3^2 N^2 \sin \left ( \pi {k\over N}
 \right ) \sim \lambda_2 N \sin \left ( \pi {k\over N}
 \right )\,.
\end{equation}

\section{Summary and discussion}

In this paper we used the Eguchi-Kawai large-$N$ volume reduction to calculate the $k$-string tension in 3d gauge theories with adjoint fermions. As we have elaborated in the previous section, our derivation relies on the limit $\lambda_3 RN \gg 1$. In this limit the W-bosons masses go to zero and hence our assumption that the dynamics is governed by a $U(1)^N$ theory may not be justified. On the other hand, we cannot exclude the option that for dynamical reasons the 3d string tension {\it is} dominated by the contribution of the Cartan sub-algebra, as in various other models \cite{Seiberg:1994rs}.   

Our work can be extended in various directions: it is natural to start from a 4d theory with adjoint fermions and to consider it on $R^2 \times T^2$ and to calculate the string tension in 4d by using the reduced 2d theory. Other directions would be to calculate the quark condensate and the glueball masses by the reduced theory.
 
The reduction from 4d to 2d raises several conceptual problems. In particular the understanding of the 4d running of the gauge coupling from the 2d theory. Once the running of the coupling is understood, it will be possible to discuss issues such as asymptotic freedom and the range of the conformal window in 4d from 2d. We hope to return to these problems in a future work. 

\vspace{0.5cm}

{\it \bf Acknowledgements.} We wish to thank A. Cherman, M. Unsal, J. Wosiek and L. Yaffe for useful discussions. 
Part of this work was done during the ``large-$N$ gauge theories" program at the GGI in Florence. 
A.A. and D.D. wish to thank the Coll\`ege de France for its warm hospitality while part of this work was carried out.

\appendix
\section{Majorana Fermions}

In this appendix we will briefly discuss the case in which the starting $3$ dimensional theory contains Majorana rather than Dirac fermions. This discussion makes possible the analysis of the ${\mathcal{N}}=1$ supersymmetric case which after reduction will give us a ${\mathcal{N}}=(1,1)$ 
SUSY theory in $2$ dimensions.

The Dirac matrices are constructed using Pauli $\sigma$ matrices: $\gamma^0=\sigma_1,\,\gamma^1=i\sigma_2,\,\gamma^2=i\sigma_3$ and satisfy
the usual Clifford algebra with Minkowski signature. The Majorana condition reads $\Psi^\ast= \sigma_3 \Psi$, so denoting the two components of $\Psi$ as $\Psi_R$ and $\Psi_L$
(with a slight abuse of notation since clearly in $3d$ there is no chirality) we have simply
$\Psi_R=\Psi_R^\ast$ while $\Psi_L=-\Psi_L^\ast$.
Since our fermions are in the adjoint representation of the gauge group it is consistent to impose a Majorana constraint on them and after
the reduction to $2$ dimensions for every $3d$ Majorana we get $2$ Majorana-Weyl fermions.

The reduction of the action is straightforward and using the above conditions it is easy to obtain:
\bea
& & \mu \exp (i \psi _{mn}) = \Psi^{\ast}_{L\, nm} \Psi_{R\, mn} =\Psi^{\ast}_{R\, mn}\Psi_{L\,nm}= \mu \exp (-i \psi _{nm})\,,\label{MajCase}
\eea
which gives us directly the antisymmetricity of the phases $\psi_{mn}$ (modulo $2\pi$ factors).

As a consequence of eq.(\ref{MajCase}) the effective term for the anomaly can be simplified to
\beq
 \delta {\mathcal{L}}_{eff}=\sum_n  F_n \left[ \frac{-i}{4\pi}\sum_{m} \ln \left(
\frac{\Psi^{\ast}_{L\,nm}\Psi_{R\,mn}}{\Psi^{\ast}_{L\, mn}\Psi_{R\,nm}} \right) \right]\nonumber  \,.
\eeq
Using further the Majorana condition we can rearrange the Yukawa term in the form
\begin{equation*}
 i g \mbox{Tr} \left( [\phi,\bar{\Psi}]\gamma^3\Psi\right) = 2 g \mbox{Tr} \left( [\phi,\Psi_L] \Psi_R\right)\,,
\end{equation*}
and after rewriting this term together with the gauge iteractions and the effective anomaly in terms of
the phases $\psi_{mn}$ we can reproduce precisely eq.(\ref{Seff}) in the Majorana case as well.

Nothing new happens when we try to solve the equation of motions, we can directly substitute the solution presented in eq.(\ref{SolFirst})-(\ref{SolLast})
obtaining for the string tension:
\begin{equation}
 \sigma_k \sim  \frac{m\mu}{2} N \sin \left ( \pi {k\over N}
 \right )\,,
\end{equation}
with an the extra factor $1/2$ with respect to the Dirac case. 



\begin{thebibliography}{99}

\bibitem{'tHooft:1973jz}
  G.~'t Hooft,
  ``A Planar Diagram Theory for Strong Interactions,''
  Nucl.\ Phys.\  {\bf B72}, 461 (1974).

\bibitem{Eguchi:1982nm}
  T.~Eguchi, H.~Kawai,
  ``Reduction of Dynamical Degrees of Freedom in the Large N Gauge Theory,''
  Phys.\ Rev.\ Lett.\  {\bf 48}, 1063 (1982).

\bibitem{Bhanot:1982sh}
  G.~Bhanot, U.~M.~Heller, H.~Neuberger,
  ``The Quenched Eguchi-Kawai Model,''
  Phys.\ Lett.\  {\bf B113}, 47 (1982).
  

\bibitem{Kovtun:2007py}
  P.~Kovtun, M.~Unsal, L.~G.~Yaffe,
  ``Volume independence in large N(c) QCD-like gauge theories,''
  JHEP {\bf 0706}, 019 (2007).
  [hep-th/0702021 [HEP-TH]].

\bibitem{adjoint-EK}

  M.~Unsal,
  ``Phases of N = infinity QCD-like gauge theories on S**3 x S**1 and nonperturbative orbifold-orientifold equivalences,''
  Phys.\ Rev.\  {\bf D76}, 025015 (2007).
  [hep-th/0703025 [HEP-TH]];

  P.~F.~Bedaque, M.~I.~Buchoff, A.~Cherman, R.~P.~Springer,
  ``Can fermions save large N dimensional reduction?,''
  JHEP {\bf 0910}, 070 (2009).
  [arXiv:0904.0277 [hep-th]];

  G.~Cossu, M.~D'Elia,
  ``Finite size phase transitions in QCD with adjoint fermions,''
  JHEP {\bf 0907}, 048 (2009).
  [arXiv:0904.1353 [hep-lat]];

  B.~Bringoltz,
  ``Large-N volume reduction of lattice QCD with adjoint Wilson fermions at weak-coupling,''
  JHEP {\bf 0906}, 091 (2009).
  [arXiv:0905.2406 [hep-lat]];

  B.~Bringoltz, S.~R.~Sharpe,
  ``Non-perturbative volume-reduction of large-N QCD with adjoint fermions,''
  Phys.\ Rev.\  {\bf D80}, 065031 (2009).
  [arXiv:0906.3538 [hep-lat]];

  T.~J.~Hollowood, J.~C.~Myers,
  ``Finite Volume Phases of Large-N Gauge Theories with Massive Adjoint Fermions,''
  JHEP {\bf 0911}, 008 (2009).
  [arXiv:0907.3665 [hep-th]];

  B.~Bringoltz,
  ``Partial breakdown of center symmetry in large-N QCD with adjoint Wilson fermions,''
  JHEP {\bf 1001}, 069 (2010).
  [arXiv:0911.0352 [hep-lat]];

  E.~Poppitz, M.~Unsal,
  ``Comments on large-N volume independence,''
  JHEP {\bf 1001}, 098 (2010).
  [arXiv:0911.0358 [hep-th]];

  A.~Hietanen, R.~Narayanan,
  ``The large N limit of four dimensional Yang-Mills field coupled to adjoint fermions on a single site lattice,''
  JHEP {\bf 1001}, 079 (2010).
  [arXiv:0911.2449 [hep-lat]];

  E.~Poppitz, M.~Unsal,
  ``AdS/CFT and large-N volume independence,''
  Phys.\ Rev.\  {\bf D82}, 066002 (2010).
  [arXiv:1005.3519 [hep-th]];

  T.~Azeyanagi, M.~Hanada, M.~Unsal, R.~Yacoby,
  ``Large-N reduction in QCD-like theories with massive adjoint fermions,''
  Phys.\ Rev.\  {\bf D82}, 125013 (2010).
  [arXiv:1006.0717 [hep-th]];

  S.~Catterall, R.~Galvez, M.~Unsal,
  ``Realization of Center Symmetry in Two Adjoint Flavor Large-N Yang-Mills,''
  JHEP {\bf 1008}, 010 (2010).
  [arXiv:1006.2469 [hep-lat]];

  A.~Hietanen, R.~Narayanan,
  ``Large-N reduction of SU(N) Yang-Mills theory with massive adjoint overlap fermions,''
  Phys.\ Lett.\  {\bf B698}, 171-174 (2011).
  [arXiv:1011.2150 [hep-lat]].

\bibitem{Armoni:2004uu}
  A.~Armoni, M.~Shifman, G.~Veneziano,
  ``From superYang-Mills theory to QCD: Planar equivalence and its implications,''
  In *Shifman, M. (ed.) et al.: From fields to strings, vol. 1* 353-444.
  [hep-th/0403071].

\bibitem{Kovtun:2004bz}
  P.~Kovtun, M.~Unsal, L.~G.~Yaffe,
  ``Necessary and sufficient conditions for non-perturbative equivalences of large N(c) orbifold gauge theories,''
  JHEP {\bf 0507}, 008 (2005).
  [hep-th/0411177].

\bibitem{kstring}
  B.~Lucini, M.~Teper,
  ``Confining strings in SU(N) gauge theories,''
  Phys.\ Rev.\  {\bf D64}, 105019 (2001).
  [hep-lat/0107007];

  L.~Del Debbio, H.~Panagopoulos, P.~Rossi, E.~Vicari,
  ``k string tensions in SU(N) gauge theories,''
  Phys.\ Rev.\  {\bf D65}, 021501 (2002).
  [hep-th/0106185];

  A.~Armoni, M.~Shifman,
  ``Remarks on stable and quasistable k strings at large N,''
  Nucl.\ Phys.\  {\bf B671}, 67-94 (2003).
  [hep-th/0307020];

  J.~Greensite, B.~Lucini, A.~Patella,
  ``k-string tensions and the 1/N expansion,''
  [arXiv:1101.5344 [hep-th]];

  A.~Athenodorou, B.~Bringoltz, M.~Teper,
  ``Closed flux tubes and their string description in D=2+1 SU(N) gauge theories,''
  JHEP {\bf 1105}, 042 (2011).
  [arXiv:1103.5854 [hep-lat]].

\bibitem{sin}
  M.~R.~Douglas, S.~H.~Shenker,
  ``Dynamics of SU(N) supersymmetric gauge theory,''
  Nucl.\ Phys.\  {\bf B447}, 271-296 (1995).
  [hep-th/9503163].

  A.~Hanany, M.~J.~Strassler, A.~Zaffaroni,
  ``Confinement and strings in MQCD,''
  Nucl.\ Phys.\  {\bf B513}, 87-118 (1998).
  [hep-th/9707244];

  C.~P.~Herzog, I.~R.~Klebanov,
  ``On string tensions in supersymmetric SU(M) gauge theory,''
  Phys.\ Lett.\  {\bf B526}, 388-392 (2002).
  [hep-th/0111078];

  A.~Armoni, M.~Shifman,
  ``On k string tensions and domain walls in N=1 gluodynamics,''
  Nucl.\ Phys.\  {\bf B664}, 233-246 (2003).
  [hep-th/0304127].

\bibitem{Coleman:1975pw}
  S.~R.~Coleman, R.~Jackiw, L.~Susskind,
  ``Charge Shielding and Quark Confinement in the Massive Schwinger Model,''
  Annals Phys.\  {\bf 93}, 267 (1975).
  
\bibitem{Gross:1995bp}
  D.~J.~Gross, I.~R.~Klebanov, A.~V.~Matytsin, A.~V.~Smilga,
  ``Screening versus confinement in (1+1)-dimensions,''
  Nucl.\ Phys.\  {\bf B461}, 109-130 (1996).
  [hep-th/9511104].

\bibitem{Armoni:1997ki}
  A.~Armoni, Y.~Frishman, J.~Sonnenschein,
  ``The String tension in massive QCD in two-dimensions,''
  Phys.\ Rev.\ Lett.\  {\bf 80}, 430-433 (1998).
  [hep-th/9709097].


\bibitem{Witten:1978ka}
  E.~Witten,
  ``Theta Vacua In Two-dimensional Quantum Chromodynamics,''
  Nuovo Cim.\  {\bf A51}, 325 (1979).

\bibitem{QCDLeff} 

  C.~Rosenzweig, J.~Schechter, C.~G.~Trahern,
  ``Is the Effective Lagrangian for QCD a Sigma Model?,''
  Phys.\ Rev.\  {\bf D21 } (1980)  3388.

  P.~Di Vecchia, G.~Veneziano,
  ``Chiral Dynamics in the Large n Limit,''
  Nucl.\ Phys.\  {\bf B171 } (1980)  253.


  E.~Witten,
  ``Large N Chiral Dynamics,''
  Annals Phys.\  {\bf 128 } (1980)  363.


\bibitem{Unsal:2010qh}
  M.~Unsal, L.~G.~Yaffe,
  ``Large-N volume independence in conformal and confining gauge theories,''
  JHEP {\bf 1008 } (2010)  030.
  [arXiv:1006.2101 [hep-th]].


\bibitem{Seiberg:1994rs}
  N.~Seiberg, E.~Witten,
  ``Electric - magnetic duality, monopole condensation, and confinement in N=2 supersymmetric Yang-Mills theory,''
  Nucl.\ Phys.\  {\bf B426}, 19-52 (1994).
  [hep-th/9407087].


\end{thebibliography}
\end{document}